\documentclass[twocolumn]{aastex61} \usepackage{amsmath}
\usepackage[utf8]{inputenc} \usepackage{xcolor}
 \begin{document} \shorttitle{The
primordial origin of LIGO's black holes} \title{Disentangling the potential
dark matter origin of LIGO's black holes} \author{Ryan Magee}
\affiliation{Institute for Gravitation and the Cosmos\\ Department of Physics
\\ Pennsylvania State University \\ 104 Davey Laboratory, University Park, PA
16802, USA\\} \email{rzm50@psu.edu}

\author{Chad Hanna} \affiliation{Institute for Gravitation and the Cosmos\\
Department of Physics \\ Pennsylvania State University \\ 104 Davey Laboratory,
University Park, PA 16802, USA\\} \affiliation{Department of Astronomy and
Astrophysics \\ The Pennsylvania State University \\ 525 Davey Laboratory,
University Park, PA 16802, USA}

\begin{abstract} The nature of dark matter (DM) remains one of the biggest open
questions in physics. One intriguing dark matter candidate, primordial black
holes (PBHs), has faced renewed interest following LIGO's detection of
gravitational waves from merging stellar mass black holes.  While subsequent
work has ruled out the possibility that dark matter could consist solely of
black holes similar to those that LIGO has detected with masses above
$10M_\odot$,  LIGO's connection to dark matter remains unknown. In this work we
consider a distribution of primordial black holes that accounts for all of the
dark matter, is consistent with all of LIGO's observations arising from
primordial black hole binaries, and resolves tension in previous surveys of
microlensing events in the Milky Way halo. The primordial black hole mass
distribution that we consider offers an important prediction--LIGO may detect
black holes smaller than have ever been observed with $\sim 1\%$ of the black
holes it detects having a mass less than the mass of our Sun and $\sim 10\%$
with masses in the mass-gap.  Approximately one year of operating advanced LIGO
at design sensitivity should be adequate to begin to see a hint of a primordial
black hole mass distribution. Detecting primordial black hole binaries below a
solar mass will be readily distinguishable from other known compact binary
systems, thereby providing an unambiguous observational window for advanced
LIGO to pin down the nature of dark matter. \end{abstract} 

\keywords{dark matter --- early universe --- gravitational waves --- stars:
black holes}

\section{Introduction}

Advanced LIGO's first observing run detected gravitational waves from the
mergers of two separate binary black hole systems (BBHs), GW150914
~\citep{Abbott2016a} and GW151226 \citep{Abbott2016b}. A third candidate event,
LVT151012 \citep{Abbott2016c}, was also observed. The second observing run is
currently underway and at least one more merger has been confirmed,
GW170104~\citep{PhysRevLett.118.221101}.  These detections prove the existence
of compact binary systems with component masses between $~7-35M_{\odot}$ and
demonstrate not only that they merge over a time scale less than the age of the
universe, but also that BBHs are relatively common and LIGO should expect to
continue to detect their coalescences~\citep{abbott2016rate}. LIGO's detections
are all consistent with relatively low spin black holes, and together they are
in agreement with a power-law distribution for the number density per unit mass of
black holes, $dN/dM\propto M^{-\alpha}$, with a 90\% credible interval $\alpha
= 2.3\substack{+1.3\\-1.4}$ \citep{abbott20161strun, PhysRevLett.118.221101}.
LIGO is or will be sensitive to binary black holes between $\sim .01-100
M_\odot$ at extra-galactic distances, and while the mass distribution was
calculated assuming a minimum black hole mass of $5M_\odot$, it remains unknown
over what mass range the presently observed power law will hold.

The notion that LIGO could detect gravitational waves from the merger of two
primordial black holes has existed for nearly two decades~\citep{Nakamura1997},
though interest in PBHs has been around for much longer~\citep{Zeldovich1967}.
Until 2004, the LIGO Scientific Collaboration actively searched for BBHs with
component masses below $1 M_\odot$~\citep{abbott2005search, abbott2008search}.
Then, the average range to a binary with component masses $(0.35, 0.35) M_\odot$ was
approximately 3 Mpc ~\citep{abbott2008search}. Advanced LIGO should
already be able to improve on that range by a factor of
ten~\citep{abbott2016gw150914}.  The PBH theory of dark matter fell out of
favor with stronger constraints from microlensing searches placing strict
limits on primordial black holes below a solar mass
~\citep{Lasserre2000,Alcock2000,Tisserand2006}, though the idea recently
resurfaced with the first model suggesting that all dark matter consisted of
PBH with a nearly monochromatic distribution around $\sim30
M_\odot$~\citep{Bird2016,Clesse2016}.  It has since been pointed out that the
monochromatic scenario has two problems.  First, constraints from dwarf galaxy
dynamics and radio emissions imply that not all of dark matter can be explained
by $\sim30 M_\odot$ black holes~\citep{Brandt2016,Koushiappas2017,Gaggero2016}.
Second, the expected merger rate predicted by this model would be above the
inferred merger rate provided by
LIGO~\citep{sasaki2016primordial,Eroshenko2016,PhysRevLett.118.221101}.  These
concerns suggest that if LIGO is probing a population of PBHs with its current
detections, PBHs might only make up a small fraction of the dark matter in the
universe.  We address both concerns in this letter.

\section{An overview of constraints on primordial black hole dark matter}

For a given PBH distribution, $n(M)$, we can define the cumulative mass density
as,
\begin{align} \rho(M_1,M_2) = \int_{M_1}^{M_2} M n(M) dM.  \end{align}

\begin{figure} \includegraphics[width=\columnwidth]{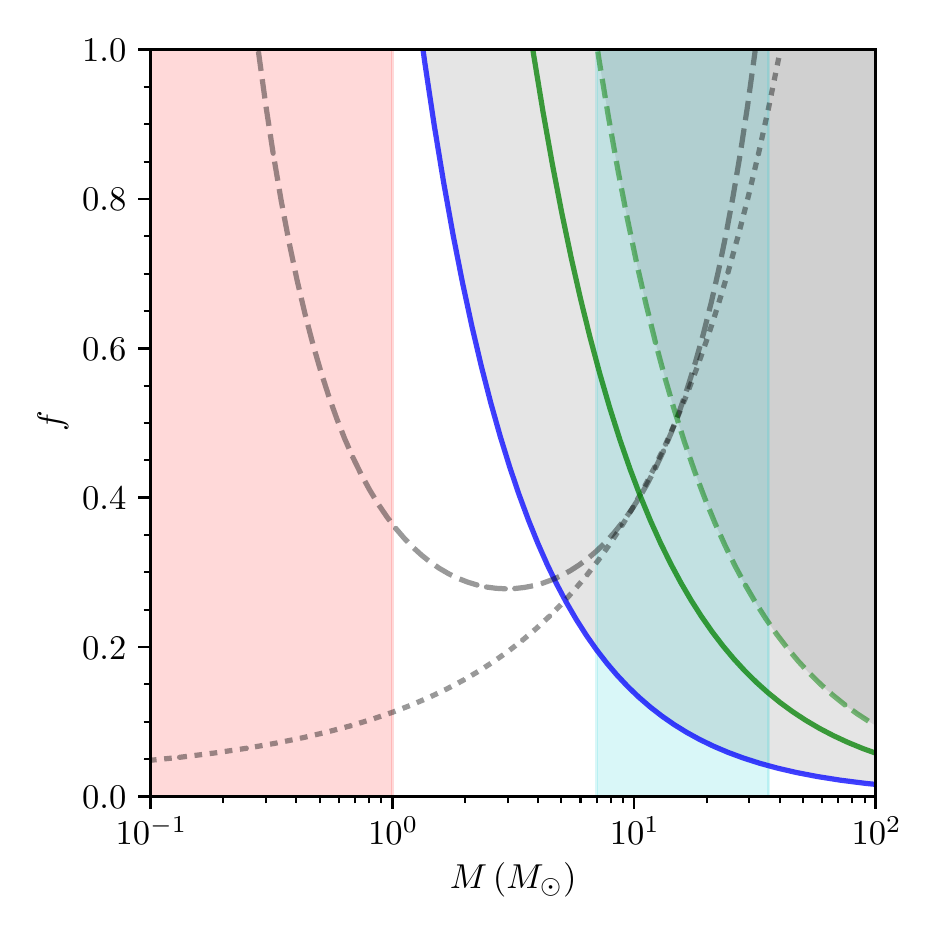}
\caption{\label{f:bounds}The maximum allowed fraction of dark matter for a
monochromatic mass function in the mass range detectable by LIGO. Section II
includes a discussion on how these may be applied to extended functions. The
red shows the range of most likely MACHO masses~\citep{Alcock2000}, while the
light blue shows the range of component masses found by LIGO. The long and
short dashed lines show the MACHO~\citep{Allsman2000} and
EROS2~\citep{Tisserand2006} collaboration constraints on compact objects for
the standard halo. The blue line shows a 99.9\% confidence limit derived from
mass segregation in Segue I ~\citep{Koushiappas2017} while the green lines are
derived from the survival of a star cluster in Eridanus
II~\citep{Brandt2016,Carr2016} for velocity dispersions and dark-matter
densities at the galactic center of $\sigma = 5,10 \text{ km }\text{s}^{-1}$
respectively and $\rho = .1M_\odot \text{pc}^{-3}$.  The shaded regions are
excluded by these constraints. Our mass function is designed to satisfy the
constraints set by the blue curve for $M>1M_\odot$. We allow it to break the
constraint set by the lower dotted line (EROS2) while mostly satisfying those
set by the MACHO collaboration since model parameters can have a large effect
on the constraints and because the MACHO collaboration concluded that there was
a population of compact objects that could account for a substantial fraction
of the dark matter. Section III discusses these justifications in more depth.}
\end{figure}

We can then write the fraction of DM made up of PBHs in some mass range as
\begin{align}f(M_1,M_2) = \frac{\rho(M_1,M_2)}{\rho_{CDM}} \end{align}
where $\rho_{CDM}$ is the presently observed cold dark matter density and
$f\leq1$. Constraints on PBH dark matter are typically presented assuming a
monochromatic distribution, but the consistency of LIGO's detections with a power
law distribution raises the question: is it possible for an extended PBH
distribution compatible with LIGO's observations to account for 100\% of the
dark matter? Constraints on monochromatic distributions can still provide
meaningful limits on extended spectra such as the one we later consider. The
constraint curves (shown in Fig.~\ref{f:bounds} for the LIGO region) bound
$df/dM$, and they also limit the cumulative fraction contained across any given
interval; for example, the EROS2 curve in Fig.~\ref{f:bounds} implies that the total
fraction of dark matter in PBHs of mass $.3M_\odot<M<3M_\odot$ is  $\lesssim
30\%$ and in masses $M>\sim 3M_\odot$ is $\lesssim 30 \%$. ~\cite{Carr2016}
points out that these two tests are not enough to guarantee the compliance of a
given extended mass function with the constraints, however.  In general, this
is a difficult task and the most rigorous test is to explicitly recalculate
every constraint for a given distribution, which several groups have already
examined for various functional forms. The primary constraints that we consider
in this work are already in a cumulative form that limits the fraction of dark
matter that can be composed of primordial black holes above a given mass.  

Quasi log-normal forms have been extensively analyzed by~\cite{Green2016}
and~\cite{Kuhnel2017}, with the former ruling out this family of mass functions
as a source of all DM for $10^{-7}M_{\odot}<M<10^{5}M_{\odot}$ while the latter
showed that for a narrow window where $10^{-10}M_{\odot}<M<10^{-8}M_{\odot}$
this remains a possibility.  These results all assume some of the strictest
interpretations of the microlensing constraints.

In the LIGO region, microlensing
~\citep{Alcock1996,Alcock2000,Allsman2000,Lasserre2000,Tisserand2006,Wyrzykowski2011}
and star cluster dynamics in the Eridanus II \citep{Brandt2016} and Segue I
dwarf galaxies \citep{Koushiappas2017} place the tightest constraints on PBH
dark matter (Fig.~\ref{f:bounds}). If the Eri II and Segue I constraints are relaxed,
power-law, log-normal, and critical collapse mass functions could allow for
100\% PBH dark matter for two small windows surrounding $5\times
10^{-16}M_\odot$ and $2\times 10^{-14}M_\odot$ as well as for $~25-100M_\odot$,
though Carr finds that no more than 10\% can be accounted for by these mass
functions when the dwarf galaxy constraints are considered~\citep{Carr2017}.
The first two windows (as well as the one~\cite{Kuhnel2017} describe) are
unreachable by LIGO, and the $\sim25-100M_\odot$ interval could yield LIGO
detection rates inconsistent with the works of ~\cite{sasaki2016primordial}
and~\cite{Eroshenko2016}.  Additionally, new constraints by~\cite{Gaggero2016}
place similar limits to Eri II and Segue I on PBH DM in the window
$25-100M_\odot$ using observations with independent systematics, making this
window less likely to contain all of the dark matter.

We explore an extended PBH mass function with the following properties.  First,
it will be single-peaked.  Second, it will have a functional form and predicted
rate consistent with LIGO observations to date.  Third, it will not violate the
tightest constraints on primordial black hole binaries in the mass range
already observed by LIGO~\citep{Koushiappas2017,Brandt2016} or the microlensing
constraints set by~\citep{Alcock2000}, but it will violate those set by the
EROS2 collaboration ~\citep{Tisserand2006}. Finally, our proposed mass function
will account for 100\% of the dark matter.  While we do not explicitly
recalculate the constraints for our test function, we use our results to
generate predictions that will be directly testable by LIGO within the next few
years.  The monochromatic EROS constraints shown in Fig.~\ref{f:allowed} for
the mass range $.1 M_\odot<M<7 M_\odot$ imply PBH can account for at most $\sim
25\%$ of the dark matter, while our models place $\sim 60-80 \%$ in that range.
We note, however, that  microlensing constraints have recently been called into
question since there are several possibilities for systematic errors that would
make them far less constraining in the region of interest for this
work~\citep{Hawkins2015}. 
 
\section{Tension in the microlensing regime}

~\cite{Paczynski1986} first proposed that massive compact halo objects (MACHOs)
could be detected by searching for evidence of stellar microlensing in the
Large Magellanic Cloud (LMC).  Since then, the MACHO, EROS, and OGLE
collaborations have all sought to identify microlensing candidates within the
galaxy through multi-year observations of millions of stars. These
collaborations were able to place tight constraints on the amount of compact
object dark matter for $10^{-7}M_\odot<M<10M_\odot$, though the limits depend
strongly on galactic halo models and small variations in the model parameters
can have a large effect on the derived constraints. Many of these parameters
remain poorly constrained and have changed since the initial microlensing
results; for example, the rotational speed of the Milky Way at $50 \text{kpc}$
was formerly assumed to be $\sim 200 \text{km/s}$, whereas current observations
suggest a speed of $136\pm56\text{km/s}$~\citep{Sofue2013} and
$178\pm17\text{km/2}$~\citep{Bhattacharjee2014} at $53\pm15 \text{kpc}$ and
$\sim50 \text{kpc}$ respectively. Similarly, stellar velocity dispersion
observations now appear to show a rotation curve that falls off with
distance~\citep{Deason2012} rather than remaining flat as was previously
assumed, and~\cite{Bratek2012} showed that determining the mass of the Milky
Way is strongly dependent on the assumed galaxy model.  These studies show that
the accepted values of important astrophysical parameters have changed and that
these variations can affect predictions of other halo parameters. 

Any deviation from the standard halo model has profound impacts on the dark matter
constraints, which is abundantly clear from the MACHO first year
~\citep{Alcock1996} and 5.7 year results ~\citep{Alcock2000}, which were unable
to rule out a 100\% compact object halo at 90\% confidence for certain models.
~\cite{Hawkins2015} acknowledges that the models that did allow for 100\% MACHO
dark matter may not be an accurate depiction of the Milky Way, but suggests
that ruling out MACHO dark matter is premature. ~\cite{Green2017} also
explicitly explored this model dependency for the tighter EROS2 constraints and
tracked changes resulting from the variation of several parameters, confirming
Hawkins' claim that current constraints are extremely model dependent.

There is another scenario that would loosen constraints in the microlensing
region independently from the halo model. As pointed out by~\cite{Clesse2016},
microlensing surveys would be far less constraining if PBHs clustered into
sub-halos, causing spikes in the local dark matter density. If PBHs are
clustered, then the likelihood of one passing directly in front of a star under
observation is much lower than for free populations. Clusters could be detected
by microlensing surveys through caustic crossings, and there may already be
evidence for this; the MACHO collaboration identified several candidate binary
microlensing events with mostly sub-solar lens
masses~\citep{Alcock1998,Alcock1999}.

The MACHO collaboration claimed positive detections of compact halo objects
with most likely masses of $.15-.9M_\odot$~\citep{Alcock2000}, and recent
microlensing surveys have hinted at similar results~\citep{Lee2015}.  If we
interpret even one detection as a possible dark matter candidate, then we need
a distribution that spans the MACHO and LIGO mass ranges. LIGO will be
sensitive to binaries in the MACHO range, and will provide a way of making a
definitive detection or establishing constraints with completely independent
systematics from those used to derive current limits on PBH DM.

\section{An extended primordial black hole mass function} Motivated by the LIGO
and MACHO observations, we consider a modified Schechter function for the PBH
mass distribution given by,
\begin{align} n(M) &= C \left(\frac{M^*}{M}\right)^{\alpha}e^{-\frac{M^*}{M}},
\end{align}
where the constant $C$ has units of $\text{vol}^{-1}\text{mass}^{-1}$.

This implies that the differential mass fraction can be written as,
\begin{align} \frac{df}{dM} &= \frac{M}{\rho_{\text{CDM}}}n(M) &=
\frac{CM^*}{\rho_{\text{CDM}}}\left(\frac{M^*}{M}\right)^{\alpha-1}e^{-\frac{M^*}{M}}
, \label{eq:dfdm} \end{align}
which is also a modified Schechter function. We stress that we are not
suggesting any connection to the Press-Schechter formalism with our choice of
this functional form. 

We choose this distribution because it has a power-law tail that fits LIGO
observations and exhibits a natural exponential cutoff at mass $M^*$. This
allows us to place the mass function's peak within the bounds set by the MACHO
collaboration. The only additional free parameter is the overall normalization,
which we constrain by considering normalizations which integrate to the overall
mass density of dark matter so that we might account for all of dark matter
with this population of black holes.

\begin{figure} \includegraphics[width=\columnwidth]{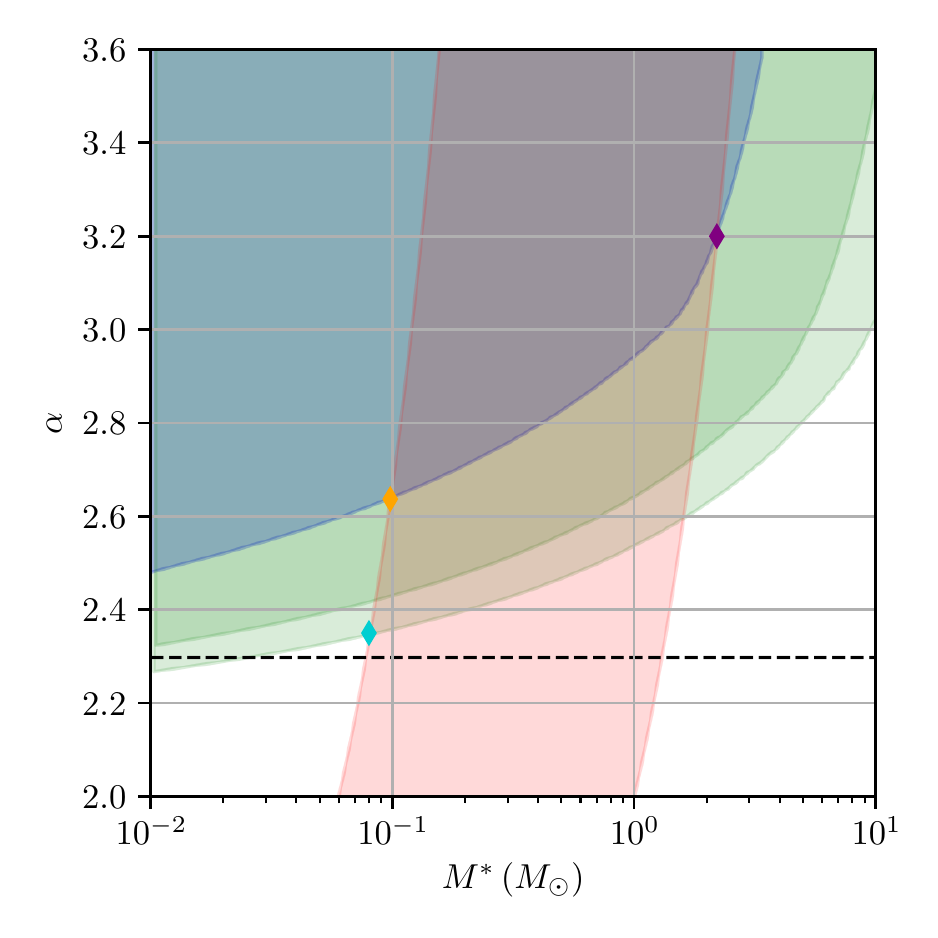} \caption{
\label{f:allowed} Allowed values of $\alpha$ and $M^*$ for the PBH distribution
\eqref{eq:dfdm}. Within the LIGO allowed values of $\alpha$, we consider the
tightest constraints on LIGO binaries comprised of PBHs given by
~\cite{Koushiappas2017}, which permits the blue region.  The green regions are
allowed by the Eridanus II constraints for the dispersions and density
referenced in Fig.~\ref{f:bounds}. The red region shows values of $\alpha$ and
$M^*$ that correspond to a mass function that peaks inside the 90\% confidence
region of the MACHO collaboration's detections across several different models
at masses of $\sim0.06-1.0M_\odot$. The purple and orange diamonds indicate two
test points, $M^* = 0.10,2.25$ and $\alpha = 2.65,3.25$ on which we evaluate
the PBH mass function to determine the dark matter distribution and the
expected LIGO rate shown in Table~\ref{table:rates}. Finally, the dotted line
shows the value of $\alpha$ expected by LIGO, and we include an additional test
point (turquoise diamond) at $M^* = .08 M_\odot$ and $\alpha=2.35$ because it
is both minimally consistent with the Eri II constraints and lies closest to
LIGO's most probable value for $\alpha$.} 

\end{figure} \begin{figure} \includegraphics[width=\columnwidth]{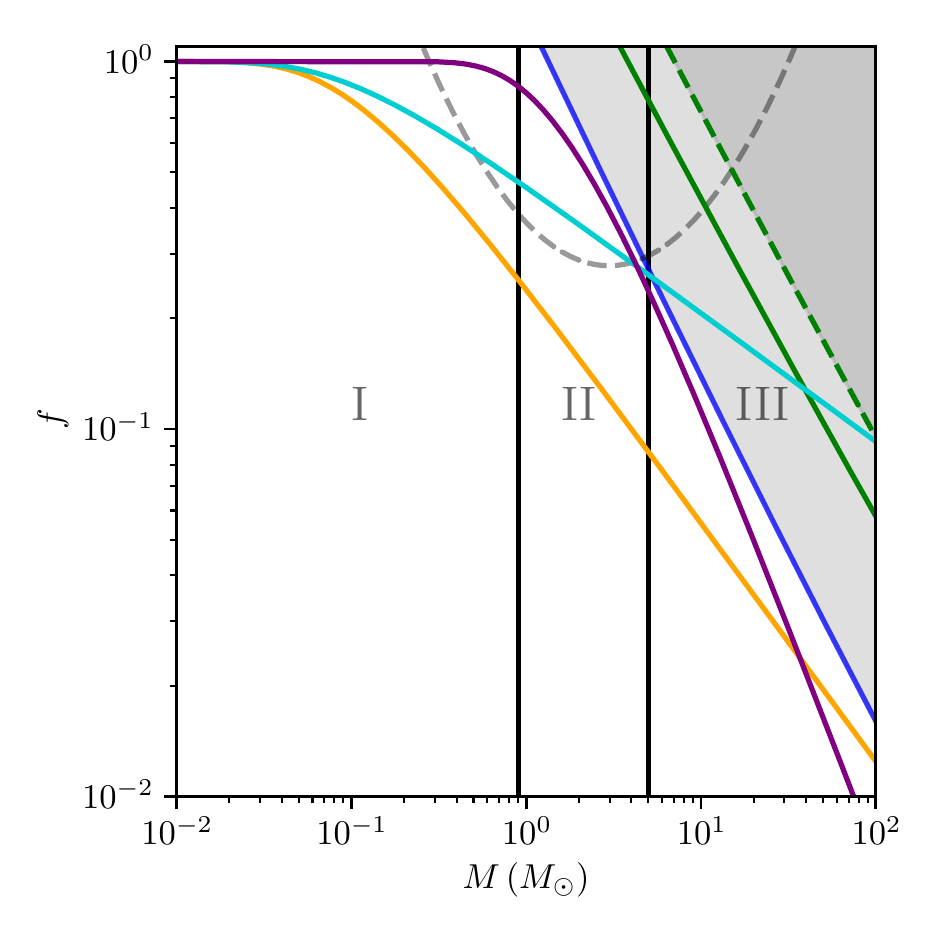}
\caption{ \label{f:integral} Satisfying constraints in the LIGO region. The
blue, green, and gray lines show the Segue I, Eridanus II, and MACHO
constraints once more.  The shaded regions are those excluded by dwarf galaxy
constraints. The purple, turquoise, and orange lines correspond to the diamonds
of the same color in Fig.~\ref{f:allowed} and show the cumulative fraction of
dark matter taken from the right. The vertical lines separate three regions of
interest: sub-solar (I), BNS and mass gap (II), and BBH region (III). Of these
observational windows, the sub-solar region would provide the most compelling
evidence for PBH given the lack of alternative theories that could produce
binary black holes at this mass. The sub-solar mass detection rate might be
$\sim .3-1\%$ of the present LIGO BBH rate providing a definitive test for the
PBH dark matter hypothesis.} \end{figure}

In Fig.~\ref{f:allowed} we show the $M^*$ and $\alpha$ values that are
consistent with LIGO observations and the dwarf galaxy
constraints\footnote{Strictly speaking, we only mandate consistency up to $100
M_{\odot}$. Though our distributions violate these constraints at higher
masses, the excess is so small ($\sim 1\%$) that uncertainties in the dark
matter density itself would need to be taken into account.}, as well as an
interpretation of the MACHO results which allows for a peak in the mass
function with objects of masses between 0.06--1$M_\odot$\footnote{We use the
fact that the function defined in \eqref{eq:dfdm} peaks at $M^* / (\alpha-1)$}.
Among the regions permitted by these constraints we favor those with the lowest
$\alpha$ to be consistent with the present most likely LIGO estimate of $\alpha
= 2.3$, though we note that all $\alpha$ in Fig.~\ref{f:allowed} are within
LIGO's 90\% confidence interval estimate.  From the highest $\alpha$ contour,
we examine the two end points $M^* = .10, 2.25 M_\odot$ and $\alpha = 2.65,
3.25$ as well as the point $M^* = .08, \alpha = 2.35$, which is the lowest
value consistent with one version of the Eridanus II constraints.  With these
parameters, we consider three observational windows defined in
Table~\ref{table:rates}. Region I includes masses up to $.9M_\odot$, which is
the lower limit on measured neutron star masses~\citep{Lattimer2012}.  Region
II contains binaries that could plausibly be neutron stars or objects in the
mass gap, and region III defines black holes consistent with current LIGO
observations.  The average masses, fraction of dark matter, and relative rates
of detection by LIGO for these windows and values of $M^*$ and $\alpha$ are
shown in Table ~\ref{table:rates}. The first two models allow for $\sim 99\%$
of the dark matter to be concentrated below $100 M_\odot$, while the last model
accounts for $\sim 90\%$. In both cases, the remaining dark matter is accounted
for by contributions above $100M_\odot$. 

\section{LIGO primordial black hole merger rates}

\begin{table}[ht] \centering \begin{tabular}{c c c c} \hline\hline Region &
Avg. Mass ($M_\odot$) & $f$ & Relative rate \\ \hline I & .1, .6, .1 & .75,
.12, .53 & .01, .01, .003 \\ II & 1.7, 1.9, 1.8 & .17, .62, .21 & .13, .22, .06
\\ III & 11, 9.3, 12.8 & .07, .25, .17 & 1, 1, 1 \\ \hline \end{tabular}
\caption{The average mass, dark matter fraction, and relative rates of
detection for LIGO in three regions of interest: I $[0,.9M_\odot)$, II
$[.9M_\odot,5M_\odot)$, and III $[5M_\odot,100M_\odot)$. These are arranged
from left to right for the models color coded in Fig.~\ref{f:allowed} and
Fig.~\ref{f:integral} as the orange ($M^* = .1, \alpha = 2.65$), purple
($M^*=2.25,\alpha=3.25$), and turquoise ($M^*=.08, \alpha = 2.35$) diamonds and
lines, respectively. For each model, the relative rate of detection by LIGO is
set to 1 for region III and calculated for the other regions
by~\eqref{eq:relrates}.\label{table:rates} }\end{table}

The space-time volume that LIGO is sensitive to scales approximately as
$M^{15/6}$, assuming equal mass binary systems with an average mass below $\sim
10 M_\odot$.  Therefore, the relative LIGO detection fraction $r$ can be
approximated as
\begin{align} r &=
\left(\frac{\overline{M_2}}{\overline{M_1}}\right)^{15/6}\times\frac{N(\overline{M_2})}{N(\overline{M_1})}
\label{eq:relrates} \end{align}
Using the average mass for each region of interest in Fig.~\ref{f:integral},
this suggests relative rates of $\mathcal{O}({1\%})$ and $\mathcal{O}({10\%})$
for the sub-solar and BNS plus mass gap regions, respectively. Our calculations
implicitly assume that the binary black hole population is identical to the
isolated population.

Several groups have already considered the formation of PBH binary systems and
predicted detection rates for LIGO. Though most assume a monochromatic,
$30M_\odot$ mass distribution, our distribution should yield similar results for
the surrounding mass range, ~$20-40 M_\odot$. These groups have found that
LIGO's observed rates are consistent with their results only if the fraction of
dark matter is small, $\sim.002<f<.02$
~\citep{Eroshenko2016,sasaki2016primordial}.  Both mass functions consistent
with the Segue I constraints predict a comparable fraction of $\sim .01-.03$
for this window, with a total fraction of $\sim.02-.04$ above $30M_\odot$.
Performing this same analysis using the looser Eridanus II constraints at
$\alpha = 2.35$, however, yields fractions in tension with these estimates with
$\sim .03-.04$ for $20M_\odot<M<40M_\odot$ and $\sim .14$ above $30M_\odot$.
Thus while LIGO's most probable $\alpha$ is marginally consistent with the
Eridanus II constraints for our distribution, it appears to violate theoretical
formation and merger rate calculations. Consequently we favor the $M^* = .1$,
$\alpha = 2.65$ distribution since it has the lowest value of $\alpha$ that
satisfies the Segue I constraints and contains the smallest fraction at or
above $30M_\odot$.

Directly applying the restrictions set by ~\cite{Eroshenko2016} and
~\cite{sasaki2016primordial} yields comparable results. Mandating that no more
than 2\% of the DM can exist at and above $30M_\odot$ predicts $\alpha \in
[2.7,3.4]$; this does not substantially change our rate predictions.  If we
instead demand that no more than $\sim.002$ exist above $30 M_\odot$, our model
predicts $\alpha \simeq 3.75$. This falls outside of the LIGO 90\% confidence
region for $\alpha$ and introduces an additional source of tension, suggesting
that a distribution of that form would be extremely unlikely.

At the end of its first observing run, advanced LIGO predicted a BBH merger
rate of $9-240\,\text{Gpc}^{-1}\text{yr}^{-1}$\citep{abbott20161strun}.  This
implies that once LIGO has surveyed 40 times the space-time volume of its first
observing run there will be a $\sim50\%$ chance of detecting 100 binary black
hole mergers. Advanced LIGO hopes to improve its reach by a factor of three,
leading to 27 times the volume accumulated per unit time. That suggests that a
year long observing run with advanced LIGO at design sensitivity could detect a
sub-solar mass binary black hole and provide smoking gun evidence for the
primordial black hole dark matter hypothesis.

In principle, LIGO should detect a binary within the BNS + mass gap region
prior to the sub-solar regime which might already hint at the origin of dark
matter. Work done by~\cite{Littenberg2015}, however, shows that few detection
scenarios provide certainty that a detected binary has a component within the mass-gap.
They found that to confidently identify component masses between $2-5M_\odot$,
hundreds of detections could be needed.  Even then, it may be difficult to
conclude that the black holes detected had primordial origins. A sub-solar
detection, however, would unambiguously provide evidence for black holes
created by non-stellar processes. We recommend a binary search in that mass
range to better study primordial black hole dark matter by systematics
independent of those used by microlensing surveys.

\section{Discussion}

Though there are many constraints on a primordial black hole model of dark
matter, there is tension in the microlensing region. We find that a modified
Schechter functional form for the differential number density can satisfy
constraints above $\sim1 M_\odot$, remain consistent with both the 90\% likely
region for several MACHO models and LIGO observations, and explain all of dark
matter. This distribution matches calculations made for the merger rate of
binary PBHs at $\sim 30 M_\odot$ and predicts a sizeable number of LIGO
detections for both the subsolar and mass gap regions, though the latter will
be difficult to distinguish from neutron stars for the foreseeable future.  In
future work, it may be worth considering the spin distribution of primordial
black hole populations. A recent study by~\cite{Chiba2017} suggests that PBHs
may be predominately slowly spinning. As more research is done into models of
PBH spin distributions and as LIGO continues to refine its measurements, PBH
spin analysis will become another strong channel for testing this model. LIGO
has the unique opportunity to place its own constraints on PBH DM abundance
while also testing for interesting new physics and we recommend searching for
binary systems below one Solar mass with LIGO in order to provide the most
definitive evidence for the PBH dark matter hypothesis. 

\section{Acknowledgments}

We thank Sarah Shandera for useful discussions. We also thank the LIGO
Scientific Collaboration and the VIRGO Collaboration, especially the Compact
Binary Coalescence working groups for their many helpful comments and
suggestions. We are grateful for the useful comments provided by the referees.
This research was supported by the National Science Foundation through
PHY-1454389.  Funding for this project was provided by the Charles E.  Kaufman
Foundation of The Pittsburgh Foundation.


\end{document}